# Collinear optical weak measurements with photonic crystals

**D. R. Solli[1], C. F. McCormick[1], S. Popescu[2], R. Y. Chiao[1], J. M. Hickmann[1,3]**

[1]*Department of Physics, University of California, Berkeley, CA 94720-7300, USA;* [2]*H. H. Wills Physics Laboratory, University of Bristol, Bristol BS8 1TL, UK;* [3]*Departamento de Física, Universidade Federal de Alagoas, Cidade Universitária, 57072-970, Maceió, AL, Brazil*
*E-mail:* hickmann@socrates.berkeley.edu

**Abstract:** We present a theoretical and experimental study of a photonic crystal based optical system in terms of weak values that map polarization states onto longitudinal spatial position and show fast and slow light behavior.
©2003 Optical Society of America

It is well known that the expectation value of a quantum mechanical operator lies within the range of its eigenvalues. However, certain measurements of a physical observable can produce values far outside this range. Aharonov, Albert, and Vaidman first described the results of such unusual kinds of measurements as "weak values" [1]. These weak values have been theoretically discussed and experimentally demonstrated in optical systems involving the angular deflections of polarized beams passing through birefringent prisms [2].

In this work, we present a theoretical and experimental study of a frequency-dependent, polarization-sensitive, collinear optical system based on a two-dimensional, birefringent photonic crystal. The group delay or time-of-flight of an optical pulse through this system is given by the eigenvalues if the system is analyzed with pure polarization basis states or eigenstates, however, unbounded values result for general superposition states. Using the theoretical framework introduced by Aharanov, Albert, and Vaidman, we show that these phenomena can be understood in terms of quantum mechanical weak measurements.

The system we consider is based on a transparent, birefringent photonic crystal which imparts a polarization-dependent phase to electromagnetic (EM) waves. The critical aspect of the photonic crystal is the large frequency-dependent birefringence in transparent spectral regions [3]. These polarization-dependent phases are functions of frequency defined as $\varphi_{TM}(\omega)$ and $\varphi_{TE}(\omega)$ where the labels TM and TE refer to transverse magnetic and transverse electric, respectively. We introduce an EM wave propagating along the y axis, normally incident on this crystal, and define the independent vertical and horizontal polarization states $|1\rangle$ and $|2\rangle$ as parallel to the z and x axes, respectively. The crystal is free to rotate in the xz-plane by an angle $\beta$ with respect to the z-axis.

If we postselect a particular transmitted state, the complex response of this system is given by $T(\omega,\beta)=\langle\psi_f|\exp[i\Gamma(\omega,\beta)]|\psi_{in}\rangle$, where $|\psi_{in}\rangle$ and $|\psi_f\rangle$ are normalized vectors describing the polarizations of the incident and detected fields, respectively, and $\exp[i\Gamma(\omega,\beta)]$ is a complex operator which describes the action of the birefringent medium on the initial state. Since we are dealing with two-dimensional polarization rotations, we can construct this operator in terms of the usual Pauli spin matrices and I, the identity operator. For example, it is not difficult to verify that for $\beta=0$, it takes the simple form $\exp[i\Gamma(\omega,0)]=(1/2)\{\exp[i\varphi_{TE}(\omega)]\ (I+\sigma_z)+\exp[i\varphi_{TM}(\omega)](I-\sigma_z)]\}$. In Fig. 1, we display a theoretical contour plot of $\arg\{T(\omega,\beta)\}$ generated using simple linear models for the birefringence. We can see singular points at $\beta$ equal to $\pi/4$ and $3\pi/4$ and normalized frequency equal to one.

In Fig. 2, we display experimental results for the phase delay vs. frequency for the light transmitted through a photonic crystal positioned at different angles. As expected, the phase delay shows unusual effects near the singularity. By taking numerical derivatives of the phase delay with respect to frequency, we can generate experimental data for the weak values $<A_\omega>_W$. In Fig. 3 we show experimental measurements of the group delay at the half-waveplate frequency as a function of the waveplate angle. Clearly, the group delay assumes extreme values ("fast and slow light") of opposite sign on either side of $\beta=\pi/4$, and takes positive, non-degenerate values at $\beta=0$ and $\beta=\pi/2$.

In conclusion, we have described and experimentally demonstrated optical weak values that manifest as a longitudinal observable, i.e., the longitudinal position or time-of-flight of a wavepacket. The weak measurements of this observable may assume values corresponding to either superluminal propagation or slow light, with a very sharp transition between the two regimes. Although our experiments were performed using classical signals, the results apply equally well to quantum-mechanical experiments. Since the weak values in our experiment are extremely sensitive to the angular position near the singularity, a system of this type can be used to make precise measurements of angular position.

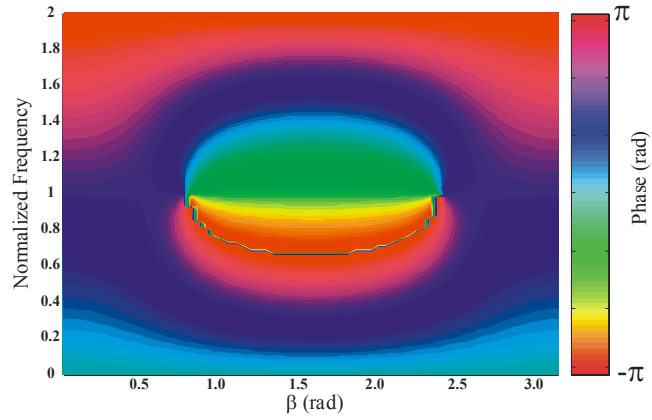

Figure 1

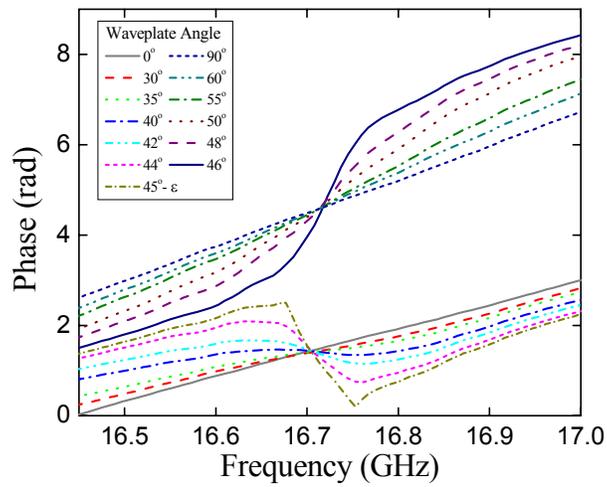

Figure 2

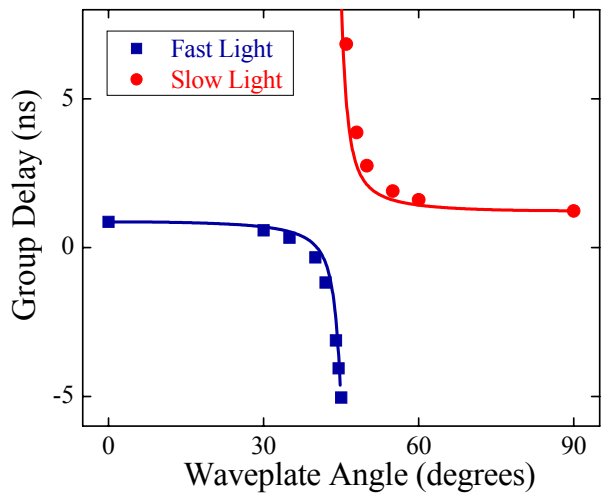

Figure 3